\pdfoutput=1
\documentclass[cits]{JINST}
\usepackage{relsize}
\usepackage{amsmath}
\usepackage{verbatim}
\usepackage{fnpos}
\usepackage{mathtools}
\usepackage{multirow}
\makeFNbottom
\newcommand*\matcalR{\includegraphics{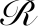}}
\title{Measurement of the front-end dead-time of the LHCb muon detector
and evaluation of its contribution to the muon detection inefficiency}
\author{
L.~Anderlini$\,^{a}$,
M.~Anelli$\,^{b}$,
F.~Archilli$\,^{c}$,
G.~Auriemma$\,^{d,\,e}$,
W.~Baldini$\,^{c,\,f}$,
G.~Bencivenni$\,^{b}$,
A.~Bizzeti$\,^{a,\,g}$,
V.~Bocci$\,^{d}$,
N.~Bondar$\,^{c,\,h}$,
W.~Bonivento$\,^{i}$,
B.~Bochin$\,^{h}$,
C.~Bozzi$\,^{c,\,f}$,
D.~Brundu$\,^{i}$,
S.~Cadeddu$\,^{i}$,
P.~Campana$\,^{b}$,
G.~Carboni$\,^{j,\,k}$,
A.~Cardini$\,^{i}$,
M.~Carletti$\,^{b}$,
L.~Casu$\,^{i}$,
A.~Chubykin$\,^{h}$,
P.~Ciambrone$\,^{b}$,
E.~Dan\'e$\,^{b}$,
P.~De Simone$\,^{b}$,
A.~Falabella$\,^{l}$,
G.~Felici$\,^{b}$,
M.~Fiore$\,^{c,\,f,\,m}$,
M.~Fontana$\,^{i}$,
P.~Fresch$\,^{d}$,
E.~Furfaro$\,^{j,\,k}$,
G.~Graziani$\,^{a}$,
A.~Kashchuk$\,^{h}$,
S.~Kotriakhova$\,^{h}$,
A.~Lai$\,^{i}$,
G.~Lanfranchi$\,^{b}$,
A.~Loi$\,^{i}$,
O.~Maev$\,^{h}$,
G.~Manca$\,^{n}$,
G.~Martellotti$\,^{d}$,
P.~Neustroev$\,^{h}$,
R.G.C.~Oldeman$\,^{i,\,o}$,
M.~Palutan$\,^{b}$,
G.~Passaleva$\,^{a}$,
G.~Penso$\,^{d,\,p}$,
D.~Pinci$\,^{d,{\mathlarger{\star}}}$, 
E.~Polycarpo$\,^{q}$,
B.~Saitta$\,^{i,\,o}$,
R.~Santacesaria$\,^{d}$,
M.~Santimaria$\,^{b}$,
E.~Santovetti$\,^{j,\,k}$,
A.~Saputi$\,^{b}$,
A.~Sarti$\,^{b,p}$,
C.~Satriano$\,^{d,\,e}$,
A.~Satta$\,^{j}$,
B.~Schmidt$\,^{c}$,
T.~Schneider$\,^{c}$,
B.~Sciascia$\,^{b}$,
A.~Sciubba$\,^{d,\,p}$,
B.G.~Siddi$\,^{f}$,
G.~Tellarini$\,^{f,\,m}$,
C.~Vacca$\,^{c,\,i}$,
R.~Vazquez-Gomez$\,^{b}$,
S.~Vecchi$\,^{f}$,
M.~Veltri$\,^{a,\,r}$
and A.~Vorobyev$\,^{h}$
\\ 
\llap{$^a$}{Sezione INFN di Firenze, Firenze, Italy} \\
\llap{$^b$}{Laboratori Nazionali dell'INFN di Frascati, Frascati, Italy} \\
\llap{$^c$}{European Organization for Nuclear Research (CERN), Geneva, Switzerland} \\
\llap{$^d$}{Sezione INFN di Roma La Sapienza, Roma, Italy} \\
\llap{$^e$}{Universit\`a della Basilicata, Potenza, Italy} \\
\llap{$^f$}{Sezione INFN di Ferrara, Ferrara, Italy} \\
\llap{$^g$}{Universit\`a di Modena e Reggio Emilia, Modena, Italy} \\
\llap{$^h$}{Petersburg Nuclear Physics Institute (PNPI), Gatchina, Russia } \\
\llap{$^i$}{Sezione INFN di Cagliari, Cagliari, Italy } \\
\llap{$^j$}{Sezione INFN di Roma Tor Vergata, Roma, Italy } \\
\llap{$^k$}{Universit\`a di Roma Tor Vergata, Roma, Italy} \\
\llap{$^l$}{CNAF-INFN, Bologna, Italy } \\
\llap{$^m$}{Universit\`a di Ferrara, Ferrara, Italy} \\
\llap{$^n$}{LAL, Universit\'e Paris-Sud, CNRS/IN2P3, Orsay, France} \\
\llap{$^o$}{Universit\`a di Cagliari, Cagliari, Italy} \\
\llap{$^p$}{Universit\`a di Roma La Sapienza, Roma, Italy} \\
\llap{$^q$}{Universidade Federal do Rio de Janeiro (UFRJ), 
Rio de Janeiro, Brazil} \\
\llap{$^r$}{Universit\`a di Urbino, Urbino, Italy} \\ \\ 

{\rule{7.5cm}{0.8pt}}  \\
$^{\mathlarger{\star}} \, \mathrm{Corresponding \; author.}$ \\ 
E-mail:\email{davide.pinci@roma1.infn.it} 
}
\abstract{
A method is described which allows to deduce the dead-time 
of the front-end electronics of the LHCb muon detector 
from a series of measurements performed 
at different luminosities at a bunch-crossing rate of 20~MHz.
The measured values of the dead-time range from $\sim 70$~ns 
to $\sim 100$~ns. These results allow to estimate the performance
of the muon detector at the future
bunch-crossing rate of 40~MHz and at higher luminosity.
}


\begin{document}
\section{Introduction}
The muon detector of the LHCb experiment \cite{jinst,muonpaper,performance}
is composed of 5 stations (M1$-$M5) placed along the beam axis.
Each station is divided in 4 regions (R1$-$R4), with increasing 
distance from the beam pipe.
Multiwire proportional chambers (MWPC) are used everywhere,
except in the most irradiated region R1 of station M1 where 
triple-GEM \cite{gem,TDR2} were adopted.
The detector comprises 1380 chambers with 122112 readout channels. 
Each chamber is segmented in anode and$\,$/or cathode elements 
named $pads$. In the front-end (FE), each pad is 
read-out by a CARIOCA chip \cite{carioca} which
performs the signal amplification, shaping and discrimination. 
In the first data taking period (years 2009-2013) the LHCb experiment
ran at a luminosity of up to $4 \times 10^{32}$ cm$^{-2}$s$^{-1}$
and at a bunch-crossing (BC) rate  $R_0 = 20$~MHz.
To investigate the performance of the muon detector at the future 
BC rate of 40~MHz and at higher luminosity, 
the knowledge of the dead-time of the CARIOCA is crucial.

The present paper is divided in two parts. In the first part
(Section~\ref{measurements}) we describe the method adopted to deduce the
dead-time of the CARIOCA ($\delta_c$) from a measure of background
rates ($R^*$) at different luminosities. These measurements were performed
in dedicated runs with a non-standard data aquisition (DAQ).

In the second part of this paper (Section~\ref{inefficiency}) 
the values of $\delta_c$
obtained in Section~\ref{measurements} are used to evaluate the muon detection
inefficiency due to the CARIOCA dead-time 
when the experiment runs in the standard 
data taking conditions and at a BC rate of 20 MHz and 40 MHz.

\section{Measurement of the CARIOCA dead time with background events}
\label{measurements}
To determine the CARIOCA dead-time, the behaviour of the FE electronics
was measured in dedicated runs at $\sqrt{s}=8$~TeV, 
at a BC rate of 20~MHz and at five different 
luminosities: $L_1 = 4 \times 10^{32}$, $L_2 = 5 \times 10^{32}$,
$L_3  = 6 \times 10^{32}$, $L_4 = 8 \times 10^{32}$
and $L_5 = 1 \times 10^{33}$ cm$^{-2}$s$^{-1}$. 
In these runs, background events were counted by  
free-running scalers placed in each FE board, 
downstream of the discriminator. These scalers are asynchronous 
with the LHC clock and can be controlled remotely. 
The dead-time of the scalers is negligible compared with
that of the CARIOCA.

The measurements were performed on the readout channels of the 
most irradiated MWPCs belonging to stations M1 and M2.
The counting rates of these channels 
are mainly due to low energy background particles. The contribution of the
electronic noise to the measured rate was evaluated without beams and 
found to be negligible. 

\subsection{Dead-time with random particles}
\label{continuous}

If the time distribution of the particles hitting a pad would not 
have any particular time structure, the counting rate ($R^*$)
of a given readout channel would be given by:
\begin{equation}
\label{delta}
R^* = R_{part}(1-\delta_c R^*)
\end{equation}
\vskip 8pt \noindent
where $\delta_c$ is the CARIOCA dead time and $R_{part}$ is the rate of 
hitting particles. 

If $R^*$ is measured at a
single luminosity, the value of $\delta_c $ cannot be 
deduced from Eq.~\ref{delta} because $R_{part}$ is unknown. 
Two measurements ($R^*_i$ and $R^*_j$) performed at two
different luminosities ($L_i$ and $L_j$) are necessary. For
each of them Eq.~\ref{delta} becomes:
\begin{equation}
\left\{
\begin{array}{ll}
R^*_i&= R^{(i)}_{part}(1-\delta_c R^*_i)\\ \\
R^*_j&= R^{(j)}_{part}(1-\delta_c R^*_j)\\
\end{array}
\right.
\label{double1}
\end{equation}
\noindent
For each pad the ratio $\rho_{ij}$, which can be evaluated from the
experimental data, is defined \cite{DP}: 
\begin{equation}
\label{ro1}
\rho_{ij}=\frac{R^*_j/L_j}{R^*_i/L_i}  \hspace{25truemm} (L_i < L_j)
\end{equation}
\vskip 8pt \noindent
Taking into account that $R_{part}$ is proportional to the luminosity
($R^{(i)}_{part}/R^{(j)}_{part} = L_i/L_j$),
Eqs.~\ref{double1} and \ref{ro1} result in the following expression:
\begin{equation}
\label{ro2}
\rho_{ij} = 1-\delta_c (1-\beta_{ij}) R^*_j  \hspace{25truemm} 
(\beta_{ij}=L_i/L_j < 1)
\end{equation}
For each pad of the detector, $\rho_{ij}$ and $R^*_j$ can be measured and 
reported on a bi-dimensional scatter plot.
According to Eq.~\ref{ro2}, the points of this plot should be aligned with 
a slope equal to $\delta_c (1-\beta_{ij})$. 
Therefore the measured value of this slope 
allows to evaluate $\delta_c$.

For particles arriving at random times, the number of hits on a pad ($N_0$)
in a time interval $\Delta t$ follows a Poisson distribution:
\begin{equation}
\label{p0}
P_{\,0}(N_0) = \frac{1}{N_0!} e^{-R_{part} \Delta t} (R_{part} \, \Delta t)^{N_0}
\end{equation}
\vskip 8pt
\noindent
\noindent and the interarrival times ($\tau$)  obey an 
exponential distribution:
\begin{equation}
\label{ptau}   
P(\tau) = R_{part} e^{-R_{part} \tau}
\end{equation}
\vskip 8pt
\noindent

With bunched particles this interarrival distribution is modified and
the distribution of the number of hitting particles in a given time interval
is no longer given by Eq.~\ref{p0}.

In the next section the real experimental situation will
be considered.

\vspace*{4truemm}

\subsection{Dead-time with bunched particles}
\label{bunched}
 
In the measurements performed at the LHC collider, $R_{part}$ has 
a bunched time structure which reflects the sequence of the
bunch crossings. Long trains of consecutive BCs occuring every 25~ns
($R_0 = 40$~MHz) or 50~ns ($R_0 = 20$~MHz) are followed by empty intervals
with no BCs. The overall duty cycle is $\sim 70$~\%.
Moreover there is a finite probability that more than one
particle generated in the same interaction will hit, near in time, the same
pad, dependently on its size and position and independently of luminosity.

A relation similar to Eq.~\ref{delta} can still be
written but $\delta_c$ must be replaced by an ``effective dead-time''
$\delta_{\,ef\!f}$~:
\begin{equation}
R^* =R_{part}(1-\delta_{\,ef\!f} R^*)
\label{eqdeltaeff}
\end{equation}
\vskip 2pt 
\noindent where $R^*$ and $R_{part}$ are the rates during the BC trains.

\noindent The value of $\delta_{\,ef\!f}$ depends on:
\begin{itemize}
\item
the CARIOCA dead-time $\delta_c$,
\item
the repetition rate of the BCs (20~MHz or 40~MHz),
\item
the probability distribution function (pdf) of the
number of particles ($N_{part}$) arriving on the pads for each BC
(the pdf changes with the luminosity),
\item
the time of arrival of particles on the pad,
\item
the time fluctuations of the chamber and FE response.
\end{itemize}

The method described in Sec.~\ref{continuous} can still be applied provided
the dependence of $\delta_{\,ef\!f}$ on the luminosity is taken into account.
Eqs.\ref{double1} becomes: 
\begin{equation}
\left\{
\begin{array}{ll}
R^*_i&= R^{(i)}_{part}(1-\delta^{(i)}_{\,ef\!f} R^*_i)\\ \\
R^*_j&= R^{(j)}_{part}(1-\delta^{(j)}_{\,ef\!f} R^*_j)\\
\end{array}
\right.
\label{double2}
\end{equation}
\noindent
By introducing these equations in $\rho_{ij}$ defined by Eq.~\ref{ro1}, 
Eq.~\ref{ro2} is replaced by the relation:
\begin{equation}
\rho_{ij} = 1- (\delta^{(j)}_{\,ef\!f} - \delta^{(i)}_{\,ef\!f}\, 
\beta_{ij})\, R^*_j  
\label{ro3}
\end{equation}

As described in the previous section, 
$\rho_{ij}$ and $R^*_j$ are measured for each pad and their values  
are reported on a
bi-dimensional scatter plot. According to Eq.~\ref{ro3} the slope of 
a linear fit to this plot is equal to 
$\delta^{(j)}_{\,ef\!f} - \delta^{(i)}_{\,ef\!f}\,\beta_{ij}$, an 
expression which therefore is computable from experimental data. 
To evaluate $\delta_c$ from this slope, 
the dependence of $\delta_{eff}$ on the luminosity and on $\delta_c$
has been estimated by a Monte Carlo (MC) simulation.

\subsection{Monte Carlo simulation of background events}
\label{MC}
The MC simulates the time of arrival of the pulses during a large 
number of consecutive BCs. 
If two events have a time distance lower than the CARIOCA
dead-time $\delta_c$\footnote{\label{note1} A gaussian fluctuation  
of $\pm 9$~ns (rms) around $\delta_c$ was assumed for each event.
Therefore $\delta_c$ represents the average CARIOCA dead-time.
This fluctuation was estimated by laboratory tests.}, 
the second one is lost.
For each BC the pdf of the number of particles hitting a pad ($N_{part}$)   
and their experimental time distribution are considered.

\subsubsection{Probability distribution function of the number of hitting
particles}
\label{231}
The number of background particles $N_{part}$ hitting a pad in a BC depends 
on the number of p-p interactions ($n_{int}$) in the BC and
on the number of particles ($n_{part}$) hitting the pad
for a single p-p interaction.

At a given luminosity, $n_{int}$ follows a Poisson statistics
with a mean value\footnote{The luminosity $L$ is proportional to $\mu$.
At $\,L = 4 \times 10^{32}$~cm$^{-2}$s$^{-1}$ and at a BC 
rate of 20 MHz, the value $\mu = 1.7$ was assumed.} $\mu$:
\begin{equation}
\label{pnint}
P_{\,1}(n_{int}) = \frac{e^{-\mu} \mu^{n_{int}}}{n_{int}!}
\end{equation}
\vskip 1pt
\noindent

For a single p-p interaction, the number of particles ($n_{part}$) crossing
a pad follows also a Poisson distribution with a mean value $\omega$: 
\vskip -8pt
\begin{equation}
\label{pnpart}
P_{\,2}(n_{part}) = \frac{e^{-\omega} \omega^{n_{part}}}{n_{part}!}
\end{equation}
\vskip 2pt
\noindent 
The value of $\omega$ represents the pad occupancy per interaction.
It is characteristic of each pad and  
depends on its size and on its position in the 
detector. The $\omega$ values are spread over several orders 
of magnitude and a maximum value of $\sim 0.05$ was measured on M2R1.

For a single BC generating an arbitrary number of interactions, the 
distribution $P_{\,3}(N_{part})$ of the number of particles 
crossing a given pad follows a ``bi-Poisson'' distribution 
which results from the convolution
of $P_{\,1}(n_{int})$ and $P_{\,2}(n_{part})$:
%
\begin{equation}
\label{pNpart}
P_{\,3}(N_{part}) = 
\begin{cases} 
\displaystyle 
\sum\limits_{n_{int} = 1}^{\infty}
\left( \frac{ e^{-\mu} \mu^{n_{int}} } {n_{int}!} \right)
\left( \frac{ e^{-n_{int}\omega}(n_{int}\omega)^{N_{part}}}
{ N_{part}! } \right)
\qquad &(N_{part} > 0)
 \\[20pt]  
\displaystyle
\sum\limits_{n_{int} = 1}^{\infty}
\left( \frac{ e^{-\mu} \mu^{n_{int}} } {n_{int}!} \right)
\left( \frac{ e^{-n_{int}\omega}(n_{int}\omega)^{N_{part}}}
{ N_{part}! } \right)
 +\, e^{-\mu}
&(N_{part} = 0)
\\
\end{cases}
\end{equation}

Because the two variables $n_{int}$ and $n_{part}$ are uncorrelated, 
the average value ($\,\overline{\vphantom{I^I}N}_{part}$) 
of $P_{\,3}(N_{part}) $ is equal to the product of the average 
values of $P_{\,1}(n_{int})$ and $P_{\,2}(n_{part})$:
\vskip 0pt
\begin{equation}
\label{Naverage}
\overline{\vphantom{I^t}N}_{part} = \omega \mu
\end{equation}
\vskip -1pt
\noindent
\noindent \vspace*{-4truemm}
and the background particle rate is:
\vskip 4pt
\begin{equation}
\label{Rpart}
R_{part} = R_0 \,\overline{\vphantom{I^t}N}_{part}  = R_0\,\mu \omega
\end{equation}
\vskip 2pt
\noindent       

In the MC, $N_{part}$ was extracted for each BC according to 
the distribution given by Eq.~\ref{pNpart}.
Note that this distribution, accounting for the probability that more
particles generated in the same interaction will hit the same pad, 
implies that, in presence of a dead time, the inefficiency 
\hbox{$(R_{part}-R^*)/R_{part} = \delta_{eff} R^*$} (see Eq.~\ref{eqdeltaeff}) 
will not go to zero for $\mu\to 0$, while $R^*$ will go to zero. 
Therefore $\delta_{eff}$, as defined
in Eq.~\ref{eqdeltaeff}, will tend to infinity for $\mu \to 0$.
This issue will be resumed later and discussed in Appendix A.

\subsubsection{Time distribution of the background hits}

The time distribution of the signals depends 
essentially on the time of flight of the detected 
particles and on the fluctuations of the response of the chamber and
its readout electronics. This distribution was extracted from 
measurements performed in special low luminosity runs where events were 
acquired in a time gate of 125~ns around the triggered BC 
(instead of 25~ns adopted in the standard data taking)
\cite{muonpaper,performance}.
Different time distributions were measured in different detector zones.

\begin{figure}[h]
\begin{center}
\vspace*{-1truemm}
\includegraphics[width=.75\textwidth]{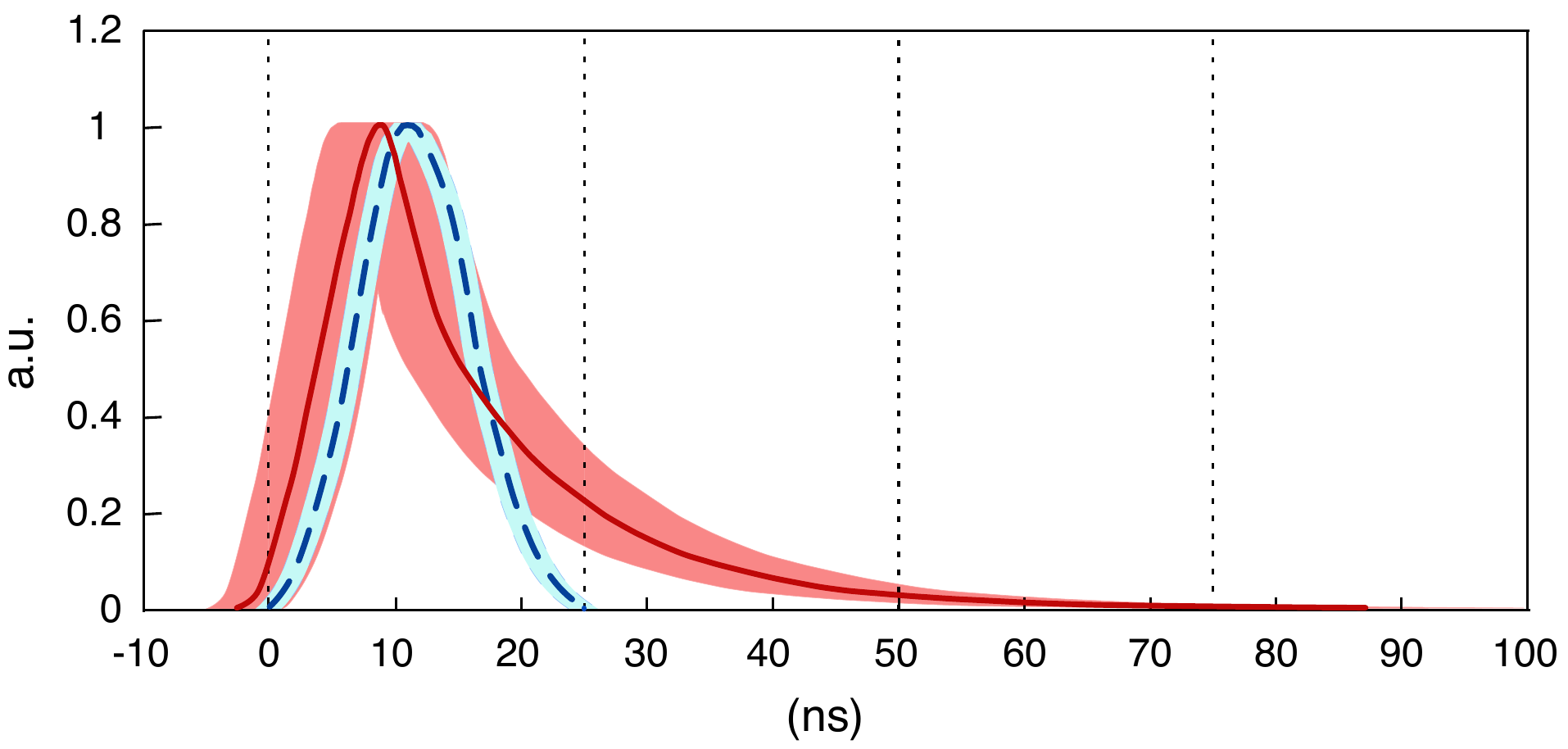}
\end{center}
\vspace*{-7truemm}
\caption{Full line: time distribution of the background hits in a single BC
used in the MC. The band around the line 
represents the variation of the distributions in the 
different chambers (see text).
Dashed line: time distribution of the muon hits.
The width of the band around the dashed line 
corresponds to a shift of $\pm 1$~ns.
The interval between two consecutive BCs can be 50 ns ($R_0 = 20$~MHz)
or 25 ns ($R_0 = 40$~MHz). 
}
\label{spectra}
\end{figure}

 In the MC, for each BC, the hits were distributed in time according to the
spectrum of Fig.~\ref{spectra} \hbox{(full line)}. The band around this line
includes the experimental time distributions measured in different chambers.
The upper and lower bounds of this band were obtained by 
stretching ($\pm 10$~\%) and shifting ($\pm 2.5$~ns) the 
full line of Fig.~\ref{spectra}. The width of the band will be relevant
to calculate the systematic errors on $\delta_c$. 

\subsection{Background Monte Carlo results}

\begin{figure}[h]
\begin{center}
\vspace*{8truemm}
\includegraphics[width=.85\textwidth]{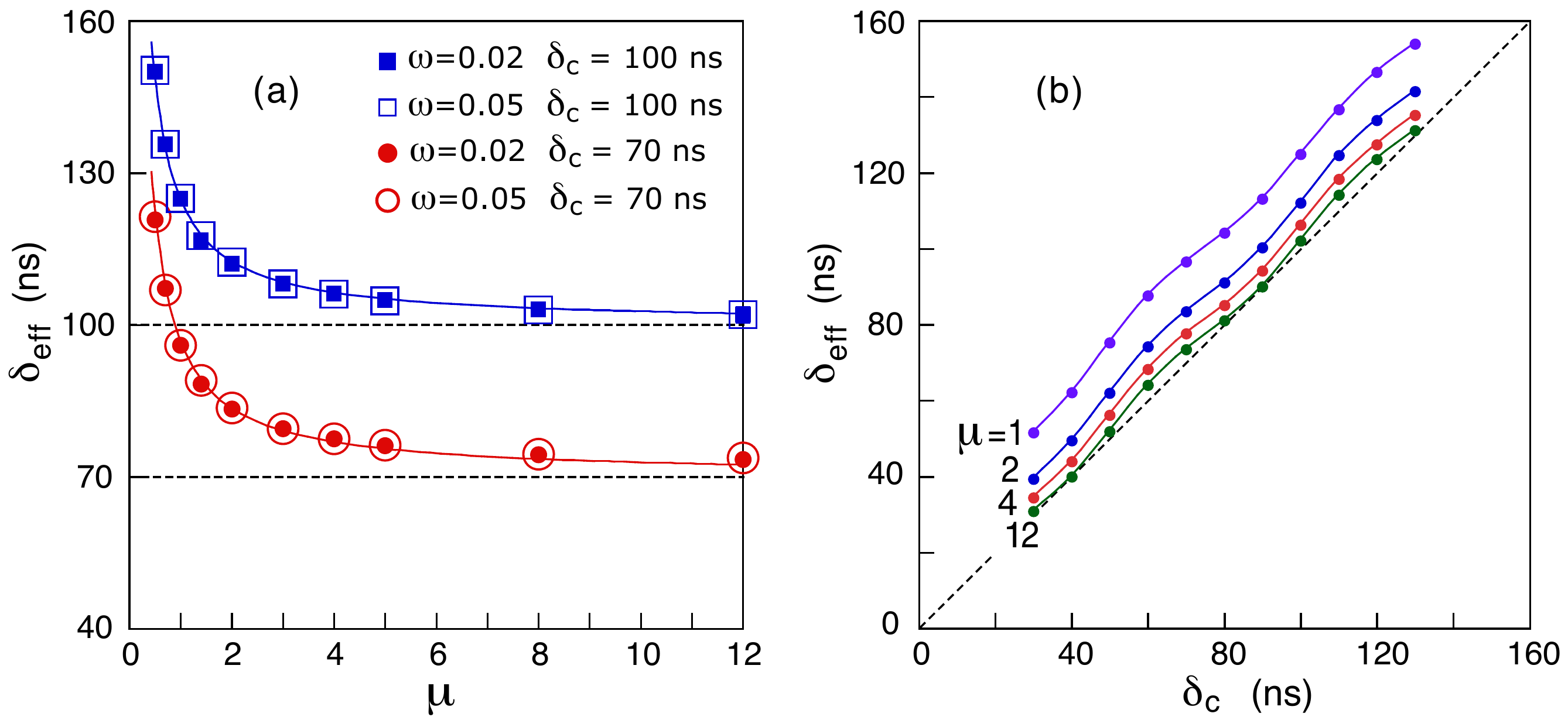}
\end{center}
\vspace*{-5truemm}
\caption{MC results: (a) Values of $\delta_{\,ef\!f}$ as a function of
$\mu$ for two values of $\omega$ and $\delta_c$.
(b) Values of $\delta_{\,ef\!f}$ as a function of $\delta_c$
for four values of $\mu$. 
}
\label{deltaeff}
\end{figure}

Once $\mu$, $\omega$ 
and $\delta_c$ are fixed, the MC determines the 
fraction of background hits which survive the
dead-time. ~The value
of $\delta_{\,ef\!f}$ is then deduced from Eq.~\ref{eqdeltaeff}.
The dependence of $\delta_{\,ef\!f}$ on $\mu$ and $\delta_c$, 
as calculated by the MC,
is shown in Figs.~\ref{deltaeff}a and \ref{deltaeff}b for
different values of $\omega$, $\delta_c$ and $\mu$.
The results reported in Fig.~\ref{deltaeff} show that the
effective dead-time increases at low luminosity (i.e. at low $\mu$)
while at high luminosity it tends to $\delta_c$\footnote{As anticipated 
in Section~\ref{231}, this behaviour is connected with the bi-poissonian
pdf of $R_{part}$. A further discussion on the matter 
is reported in Appendix A.}.
The value of $\delta_{\,ef\!f}$ does not
depend on $\omega$, being the same for all pads.
The undulation of the curves of Fig.~\ref{deltaeff}b reflects
the bunched structure of the beams.
The dependence of $\delta_{\,ef\!f}$ on $\delta_c$ and on
$\mu$ (Fig.~\ref{deltaeff}b)
allows to calculate the expression
$\delta^{(j)}_{eff} - \delta^{(i)}_{eff} \beta_{ij}\,$ and to find the 
value of $\delta_c$ for which this expression assumes its experimental
value (Eq.~\ref{ro3}).

\subsection{Experimental results}
\label{data}

The counting rates of the pads belonging to the regions of station
M1 and M2 equipped with \hbox{MWPCs} were measured in 1~s by the scalers
of the CARIOCA chip. To take into
account the duty cycle of LHC, the values of $R^*$
were obtained by attributing the scaler counts to a time interval of 0.7~s.
The measurements were performed at five different luminosities. 
Only the couples of measurements performed in the same day were 
chosen for the analysis\footnote{Larger variations are observed on data taken 
at the same nominal luminosity in different days.
This effect is probably due to the  
uncertainty on the true luminosity value.}.
The data belonging to the five selected pairs of luminosities ($L_i$, $L_j$)
were represented on a scatter plot $\rho_{ij}$ versus $R^*_j$. 

\begin{figure}[t]
\vspace*{3truemm}
\begin{center}
\includegraphics[width=.7\textwidth]{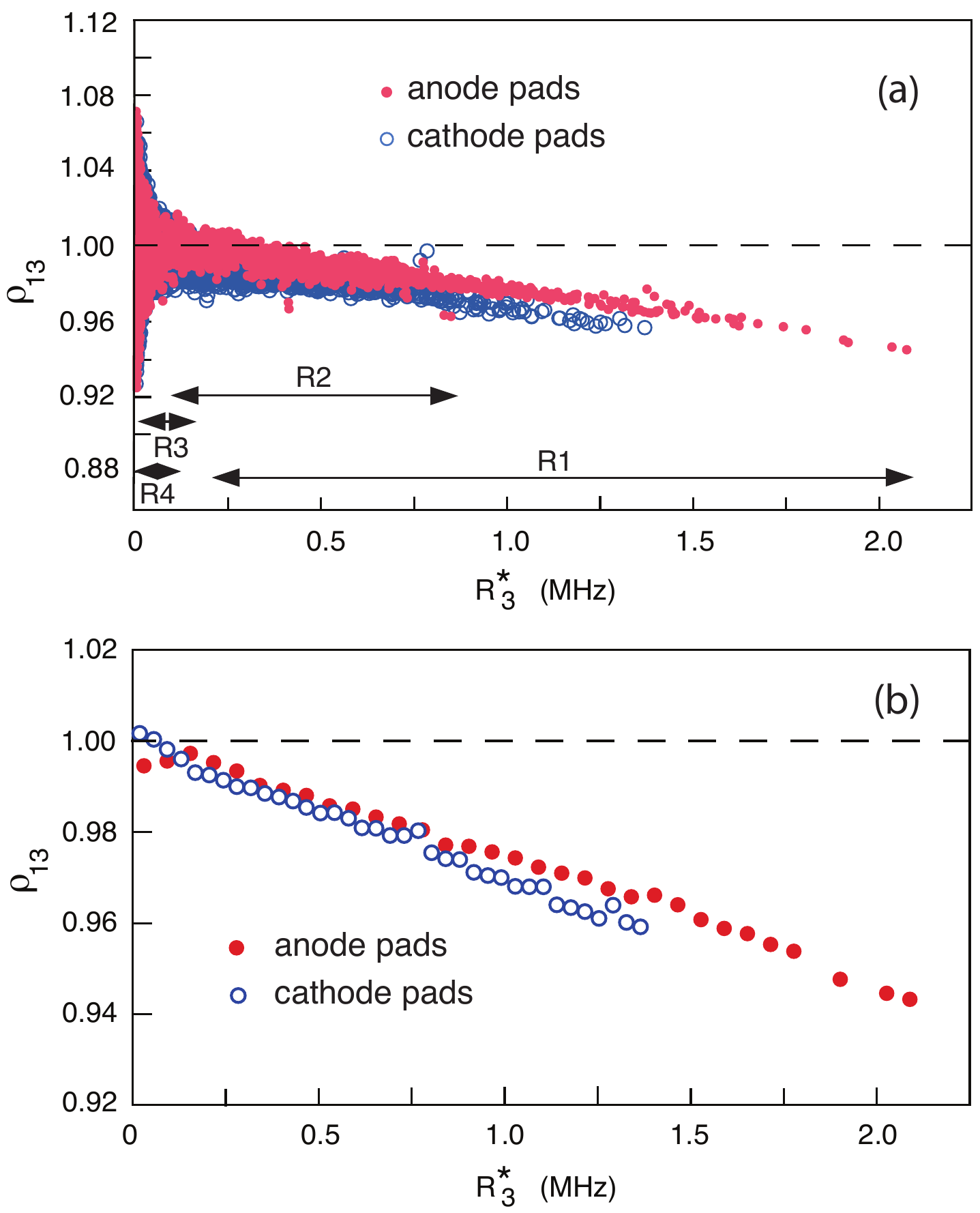}
\end{center}
\vspace*{-2truemm}
\caption{(a): Experimental values of $\rho_{13}$ 
as a function of $R^*_3$ 
(see Eq.~{\protect\ref{ro3}}).
Each point corresponds to a pad of station M2. 
Anode and cathode pads are shown separately.
The ranges of rate measured in the regions R1$-$R4 are indicated.
(b): The points reported in (a) were averaged in bins of $R^*_3$.
}
\label{scatterplotnew}
\vspace*{2truemm}
\end{figure}

As an example we show (Fig.~\ref{scatterplotnew}a) 
the results of the measurements 
performed on all the pads of station 
M2, and using the pair of luminosities ($L_1,L_3$). Each point corresponds to
a pad. Cathode and anode pads are shown separately.
The difference in the slope for cathode
and anode readout is mainly due to the CARIOCA chip which is slightly 
different for positive  and negative pulses. 
The width of the  bands is 
compatible with the expected statistical fluctuations on the number
of counted pulses. The fluctuations are larger in regions R3 and R4 
(Fig.~\ref{scatterplotnew}a) where the counting rate is lower.
The points averaged in bins of $R^*_3$,
are reported in Fig.~\ref{scatterplotnew}b. 
If no dead time effect was present, the points of 
Fig.~\ref{scatterplotnew}a and \ref{scatterplotnew}b should be 
aligned around the line $\rho_{13} = 1$. 
The experimental points are instead distributed
around a line with a non-zero slope which, according to Eq.~\ref{ro3},
is equal to $\delta^{(3)}_{\,ef\!f} - \delta^{(1)}_{\,ef\!f}\,\beta_{13}$.

\begin{figure}[t]
\begin{center}
\vspace*{0truemm}
\includegraphics[width=.6\textwidth]{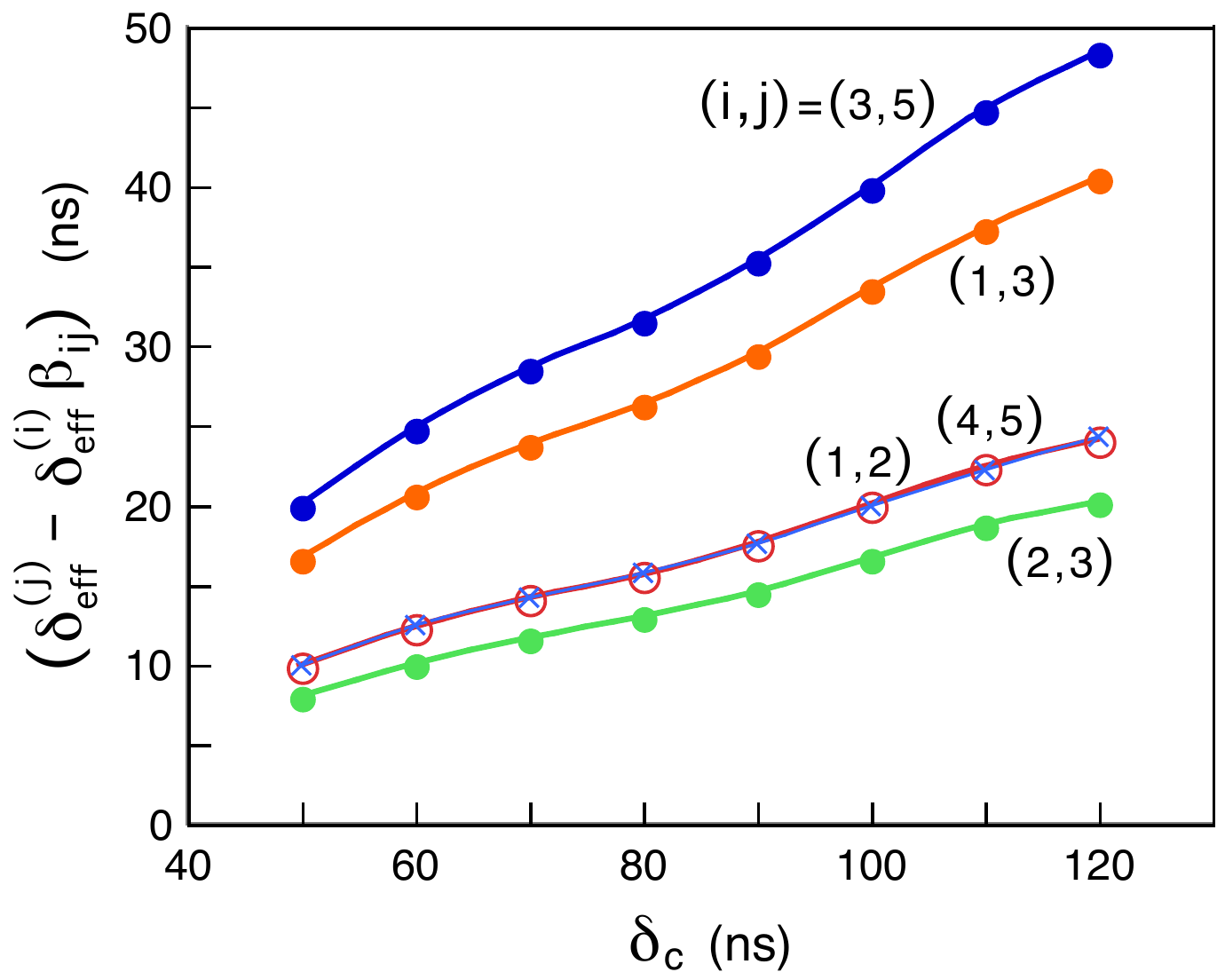}
\end{center}
\vspace*{-1truemm}
\caption{Slope of the linear dependence of $\rho_{ij}$ vs. $R^*_j$
described by Eq.~{\protect \ref{ro3}}. The curves, calculated with the MC, 
refer to the five pairs of luminosities ($L_i$, $L_j$) 
considered in the data analysis.
}
\vspace*{1truemm}
\label{slope}
\end{figure}
\vspace*{6truemm}

The procedure described for the pair of luminosities 
($L_1,L_3$) was applied to all the pairs of luminosities considered,
separately for different regions and readout.
Then the MC results shown in Fig.~\ref{slope} allow to calculate $\delta_c$
from the measured values of 
$\delta^{(j)}_{\,ef\!f} - \delta^{(i)}_{\,ef\!f}\,\beta_{ij}$.
The results are reported 
in Table~\ref{table1}, for the MWPCs of station M1 (regions R2, R3 and R4)
and M2 (regions R1 and R2), and for cathode and anode readout separately.
\begin{table}[t]
\vspace*{1truemm}
\caption{Experimental results on the dead-time $\delta_c$ (ns) of 
the FE electronics of the MWPCs belonging 
to stations M1 and M2. The results, reported in columns $3-7$,
refer to the pair of luminosities ($L_i \, , L_j$) used. 
The average value of the data 
obtained with the five pairs of luminosities is reported in the last column.
The first error is the disperion (rms) of these values. The
second error reflects the uncertainty on the time distributions (see text).
The capacitance of the pads belonging to each region and readout is reported
in column 2.
}
\label{table1}
\vspace*{-2truemm}
\begin{center}
\begin{tabular}{|l|c|c|c|c|c|c|c|}
\hline
Region \& readout & C$_{pad}$ (pF) & $\!(L_1,L_2)\!$&$\!(L_2,L_3)\!$&
$\!(L_1,L_3)\!$&$\!(L_3,L_5)\!$&$\!(L_4,L_5)\!$&average \\
\hline
M1 R2 cathode & 58 & 101 & 102 & 97 & 104  & 106 & $102 \pm 4 \pm 1$ \\
M1 R3 cathode  & 51 & 89  & 90  & 90 & 93   & 94  & $91  \pm 2 \pm 2$ \\
M1 R4 anode & 107 & 66  & 66  & 70 & 75   & 76  & $71 \pm 4 \pm 1$ \\
M2 R1 cathode & 131 & 86  & 88  & 95 & 83   & 84  & $87 \pm 4 \pm 2$ \\
M2 R1 anode & 75 & 76  & 77  & 84 & 62   & 66  & $73 \pm 7 \pm 1$ \\
M2 R2 cathode & 111 & 82  & 84  & 84 & 76   & 83  & $82 \pm 3 \pm 2$ \\
M2 R2 anode & 77 & $-$ & $-$ & $-$    & 71   & 65  & $68 \pm 3 \pm 2$ \\
\hline
\end{tabular}
\end{center}
\end{table}

For each region and readout, the $\delta_c$ values obtained from
different pairs of luminosity are significantly different. The dispersion
of these values, reported as the first error in the last column 
of Table~\ref{table1}, represents the main contribution to the systematic
error on $\delta_c$. 

The second error is due to the uncertainties
on the time distributions. To evaluate this contribution,
the time distribution of the background pulses was moved inside
the band shown in Fig.~\ref{spectra} and the maximum variation was taken. 
Also the fluctuation of the CARIOCA dead-time
around its average value $\delta_c$  was varied
in the interval $9 \pm 4$~ns (see footnote \ref{note1}). This last variation
has a small effect on the results. 

The results referring to the same pad type (anode or cathode) of 
a given station and region, but obtained with different pairs of luminosities
are in a reasonably good agreement. 
The dead-time of the cathode readout are systematically
larger than those of the anode readout. 
Considering separately the anode and cathode readout, some spread
is still present between the results belonging 
to different stations and regions.
This can be due to different detector geometries and therefore capacitances.

\begin{figure}[h]
\begin{center}
\vspace*{5truemm}
\includegraphics[width=.55\textwidth]{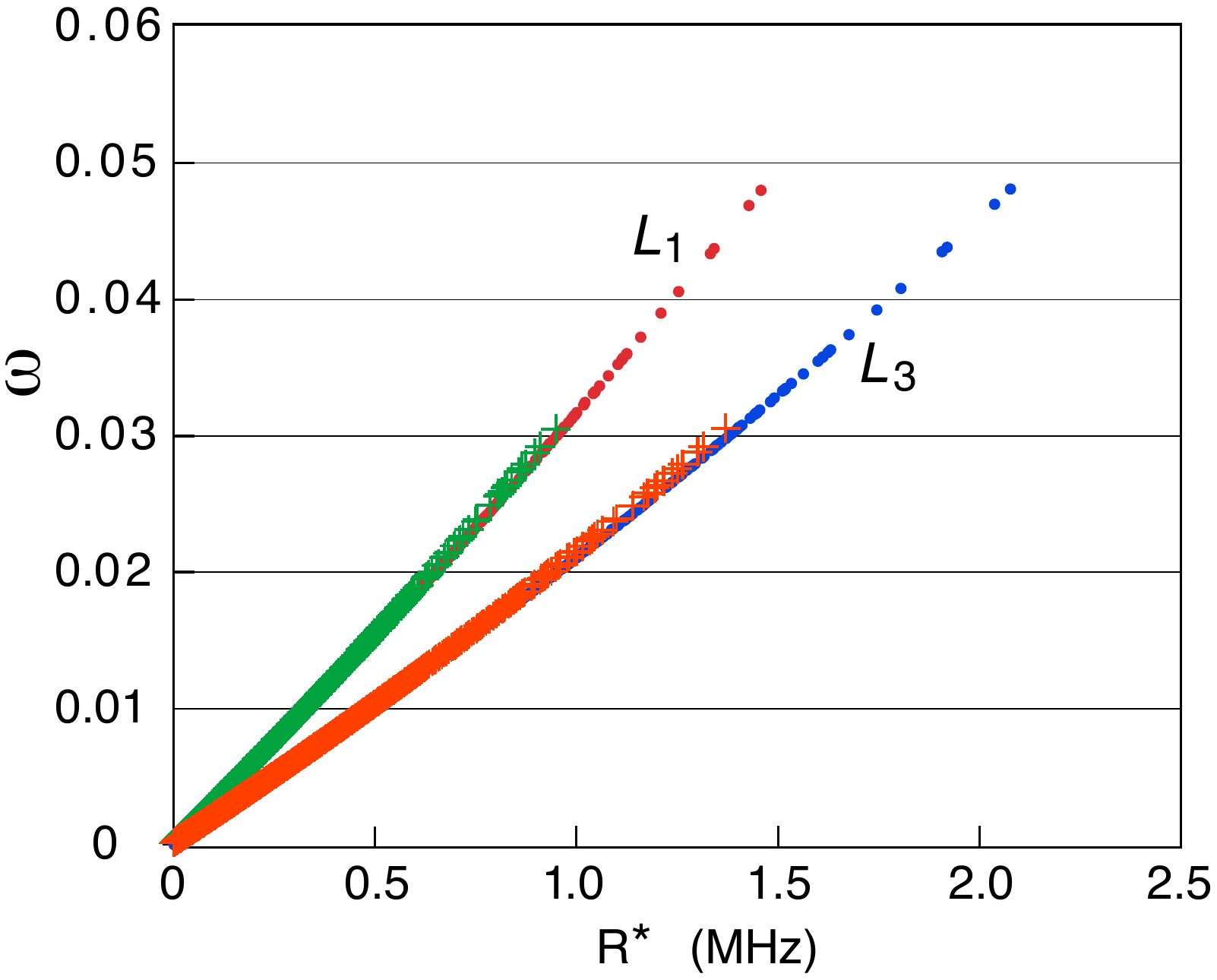}
\end{center}
\vspace*{-4truemm}
\caption{Experimental average number of background particles hitting a 
pad for one p-p interation ($\omega$) as a function 
of the pad counting rate for the two luminosities $L_1$ and $L_3$. 
Each point represents a pad of station M2: 
full points for anode pads, crosses for cathode pads. 
}
\label{omeganew}
\end{figure}

Once $\delta_c$ is determined, the values of $\delta_{eff}$ 
can be evaluated (Fig.\ref{deltaeff}b). Finally, 
Eqs.~\ref{eqdeltaeff} and \ref{Rpart}  allow to calculate, for each pad,
the corresponding value of $\omega$. As an example the dependence of 
the pad occupancy per interaction
$\omega$ on $R^*$ is  shown in Fig.~\ref{omeganew} 
separately for anode and cathode pads of station M2, for two 
luminosities $L_1$ and $L_3$.
%
\section{Front-end inefficiency for muon detection due to CARIOCA dead-time}
\label{inefficiency}
Once $\delta_c$ is determined with the described method, the detection
inefficiency for muon hits of 
each front-end\footnote{The MWPCs comprise two layers (1 gas gap per layer 
in M1 and 2 gas gaps per layer in M2). Each layer is readout by 
one one FE (one CARIOCA chip) and the two layers
are OR-ed. The inefficiecy we are speaking about is  
the one of a hit in a single layer. 
To evaluate the overall muon detection inefficiency 
of a station, the cluster size and the correlation 
of the hits in the two layers must be 
taken into account.} due to the CARIOCA dead-time can be evaluated. 
In the standard data taking the hits 
are acquired in a 25~ns gate, corresponding to the triggered BC.
If one or more particles (muon or background particles) hit the pad 
in the 25~ns gate, and at least one survives to the dead-time generated in
a preceeding BC the pad is efficient.

\rule{10pt}{0pt}\\

For a given pad the muon rate is negligible compared to the background rate
so that the muon detection inefficiency is due to the dead-time generated
by background hits. This inefficiency can be represented by an 
effective dead-time for muons ($\delta_{\,ef\!f}^{(muon)}$) defined
by the equation:

\vspace*{3truemm}
\begin{equation}
\frac{R^*_{muon}}{R_{muon}} = 1-\delta_{\,ef\!f}^{(muon)} R^*
\label{eqdeltaeffmu}
\end{equation}
\vspace*{2truemm}

\noindent where $R_{muon}$ is the true muon rate while $R^*_{muon}$ is the
measured muon rate. The first member of Eq.~\ref{eqdeltaeffmu} represents
the FE muon detection efficiency ($\varepsilon_{muon}$), 
while in the second member $R^*$ is the rate of background particles 
counted in the pad during the BC trains.
\rule{10pt}{0pt} \\

\subsection{Muon Monte Carlo simulation}

To evaluate $\delta_{\,ef\!f}^{(muon)}$ in the standard data taking 
conditions, a dedicated MC was set which simulates a muon hit occurred
in a triggered BC
superimposed to a particle background having a rate $R_{part}$. 
The fraction of lost muons
was evaluated and $\delta_{\,ef\!f}^{(muon)}$ was calculated from
Eq.~\ref{eqdeltaeffmu}.
In the triggered BC the time of the muon hit was extracted according
to the distribution \cite{muonpaper} shown 
in Fig.~\ref{spectra} (dashed line), while in all the BCs 
the background hits were extracted as described in Sec.~\ref{MC}.
\rule{10pt}{0pt} \\

\subsection{Muon Monte Carlo results}
\label{muMCresults}

\begin{figure}[b]
\begin{center}
\includegraphics[width=.90\textwidth]{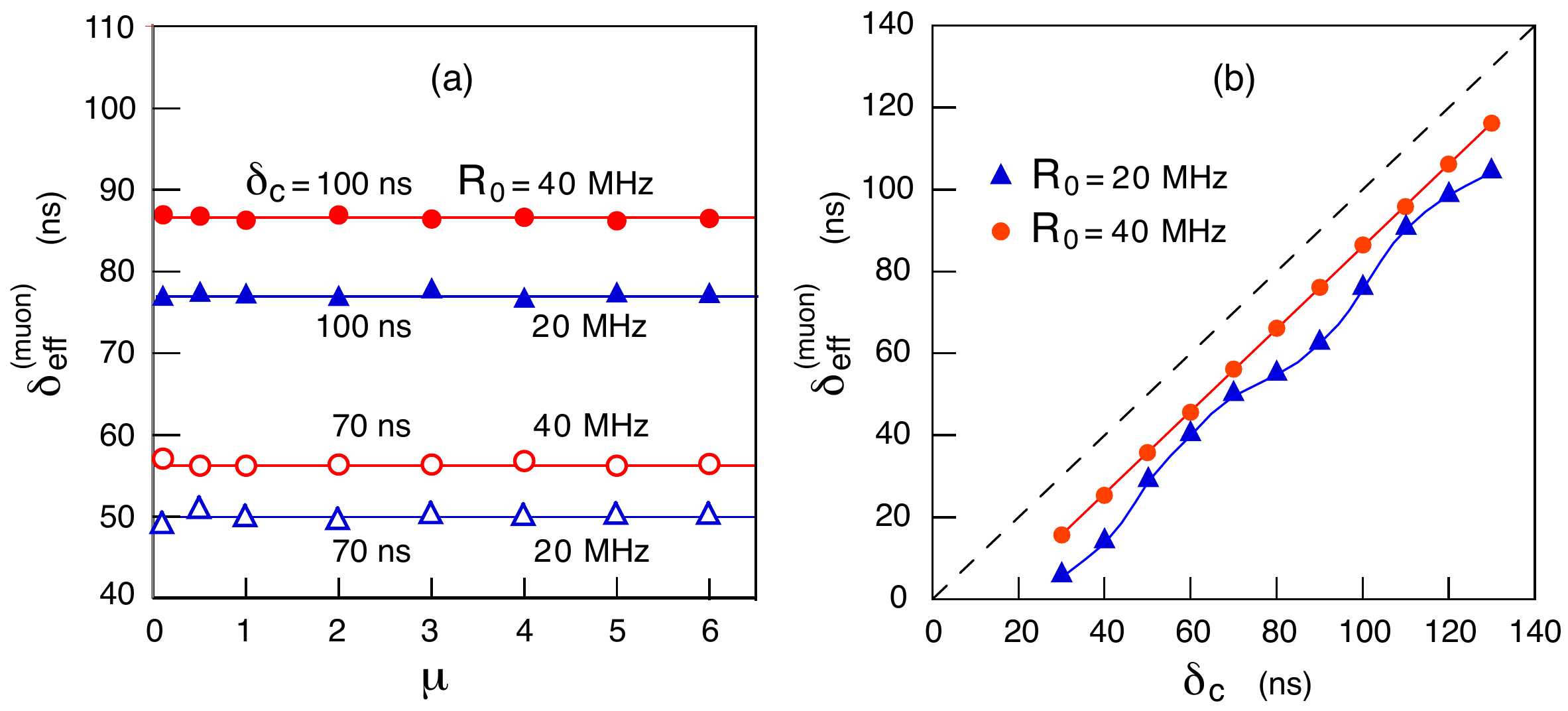} 
\end{center}
\vspace*{-6truemm}
\caption{(a) Effective dead-time for muon hit detection for two 
values of $\delta_c$ and $R_0$, as a function of
the luminosity represented by $\mu$. (b) $\delta_{\,ef\!f}^{(muon)}$
as a function of $\delta_c$ for the two values of $R_0$.
}
\label{deltaeffmuoni}
\end{figure}
The Muon MC was run for the BC rate of the LHCb Run1 (20~MHz) and for the
future BC rate of 40~MHz.
In Fig.~\ref{deltaeffmuoni}a the (in)dependence of $\delta_{\,ef\!f}^{(muon)}$
on $\mu$ is shown for two values of $\delta_c$ and $R_0$. 
In Fig.~\ref{deltaeffmuoni}b the dependence of $\delta_{\,ef\!f}^{(muon)}$
on $\delta_c$ is reported\footnote{A discussion about 
the different behaviours shown in Fig.~\ref{deltaeff}
and in Fig.~\ref{deltaeffmuoni} is reported in Appendix A.} for the two 
values of $R_0$. It allows to determine 
the values of $\delta_{\,ef\!f}^{(muon)}$ corresponding to the measured
$\delta_c$ values of Table~\ref{table1}. These values are reported 
in Table~\ref{table2}. The first error on $\delta_{\,ef\!f}^{(muon)}$
is due to the dispersion of $\delta_c$ measured at the five
different pairs of luminosity.
The second error on $\delta_{\,ef\!f}^{(muon)}$ was estimated by 
taking into account the uncertainties on the time distributions
of the background and of the muons pulses 
(bands shown in Fig.~\ref{spectra}). The large variations of the
errors  at $R_0 = 20$~MHZ,
are due to the undulation induced  on $\delta_{\,ef\!f}^{(muon)}$
by the bunched beam structure (Fig.~\ref{deltaeffmuoni}b).

\begin{table}[t]
\vspace*{-3truemm}
\caption{For each region and readout the pad area $A$ (cm$^2$) is shown
in the second column.
In the third column the average values of $\delta_c$ (ns)
of Table~{\protect\ref{table1}} are reported.
The values of $\delta_{\,ef\!f}^{(muon)}$ (ns), calculated
with the muon MC, are reported in 
the last two columns for a BC rate of 20 MHz and 40 MHz.
The first error on $\delta_{\,ef\!f}^{(muon)}$
is due to the dispersion of $\delta_c$ measured at the five
different pairs of luminosity while the second error was estimated by
taking into account the uncertaities on the time distributions
of the background and muon hits
(see text). 
}
\vspace*{2truemm}
\begin{center}
\begin{tabular}{|l|c|c|c|c|}
\hline 
\raisebox{-2.5ex}{Region \& readout} & \raisebox{-2.5ex}{$A$} & 
\raisebox{-2.5ex}{$\delta_c$} &
\multicolumn{2}{c|}{\raisebox{-1.ex}{$\delta_{\,ef\!f}^{(muon)}$}} \\ 
\cline{4-5} & & &{\raisebox{-0.2ex}{$R_0 = 20$~MHz}} & 
{\raisebox{-0.2ex}{$R_0 = 40$~MHz}} \\
\hline
M1 R2 cathode & 5 & $102 \pm 4 \pm 1$ & $79 \pm 5 \pm 5$ & $89 \pm 4 \pm 2$ \\
M1 R3 cathode & 20 & $91 \pm 2 \pm 2$  & $65 \pm 3 \pm 4$  & $77 \pm 2 \pm 2$ \\
M1 R4 anode & 80 & $71 \pm 4 \pm 1$  & $51 \pm 3 \pm 2$  & $58 \pm 3 \pm 2$ \\
M2 R1 cathode & 11.7 & $87 \pm 4 \pm 2$ & $61\pm 4 \pm 3$  & $73 \pm 3 \pm 2$ \\
M2 R1 anode & 15.6 & $73 \pm 7 \pm 1$  & $52\pm 5 \pm 1$  & $60 \pm 6 \pm 2$  \\
M2 R2 cathode & 23.5 & $82 \pm 3 \pm 2$ & $57 \pm 3 \pm 2$ & $68 \pm 3 \pm 2$ \\
M2 R2 anode & 31.25 & $68 \pm 3 \pm 2$  & $49 \pm 2 \pm 2$ & $54 \pm 2 \pm 2$ \\
\hline
\end{tabular}
\end{center}
\label{table2}
\end{table}
\vspace*{2truemm}

The muon detection efficiency of a pad is equal to
$1 -\delta_{\,ef\!f}^{(muon)} R^*$ where
$R^*$ is the background rate of the pad measured during the
duty cycle (70~\%) of LHC.
\vspace*{3truemm}\rule{10pt}{0pt} \\

\subsection{Muon hit inefficiency in the past and future running conditions}

The results reported in Table~\ref{table2} allow to assess the FE 
performance in the past runs and predict its behaviour at the high rates
foreseen in the future high-luminosity upgrade conditions.
For this purpose we express the muon 
detection efficiency  (Eq.~\ref{eqdeltaeffmu}) of a pad
as a function of the rate of background particles
($R_{part}$) hitting the pad:
\vspace*{5truemm}

\begin{equation}
\varepsilon_{muon} = \frac{R^*_{muon}}{R_{muon}} = 
1-\delta_{\,ef\!f}^{(muon)} R^*
= 1 - \delta_{\,ef\!f}^{(muon)} \frac{R_{part}}{1+\delta_{eff} \, R_{part}}
\label{effmu1}
\end{equation}
\vspace*{5truemm}

\noindent where $R_{part}$ is the particle rate on the pad during the
duty cycle (70~\%) of LHC and $\delta_{eff}$ is a 
function of $\delta_c$ (Fig.~\ref{deltaeff}b).
Taking into account the values of $\delta_c$ and $\delta_{\,ef\!f}^{(muon)}$
reported in Table~\ref{table2}, Eq.~\ref{effmu1} allows to calculate the
muon detection efficiency as a function of $R_{part}$, for the seven
considered regions and readout.
This efficiency is usually reported as a function of the particle rate
per cm$^2$ $\,\matcalR_{part} = 0.7\, R_{part}/A$, where 
$0.7\, R_{part}$ is the particle rate on the pad averaged
on the running time and $A$ is the pad area reported in Table~\ref{table2}.
\eject
\vspace*{3truemm}

\begin{figure}[t]
\begin{center}
\vspace*{5truemm}
\includegraphics[width=1.\textwidth]{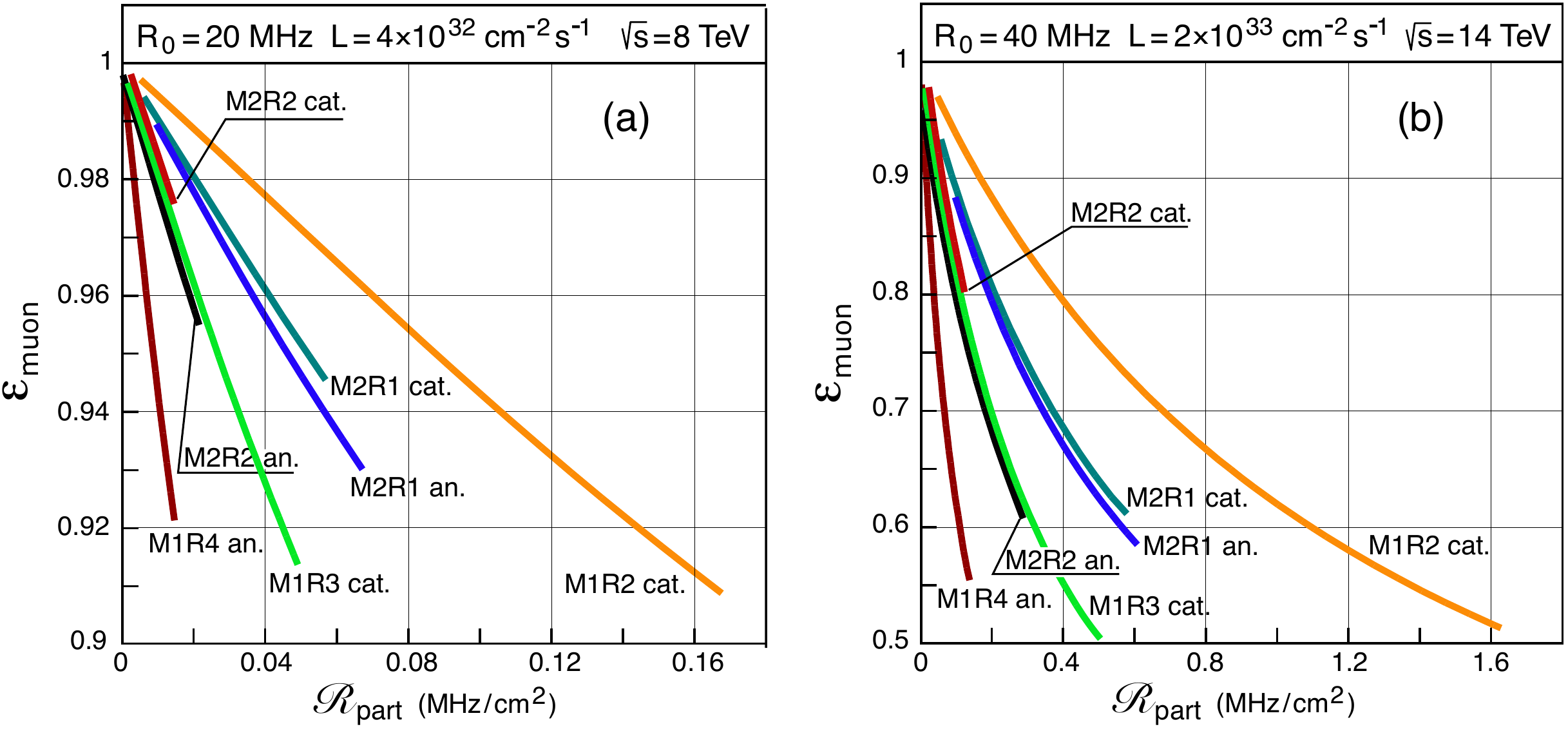}
\end{center}
\vspace*{-2truemm}
\caption{Muon detection efficiency of a single FE for the seven considered 
regions and readout (monogaps in M1 and bigaps in M2), 
as a function of the rate per unit area. 
The past experimental conditions are shown in (a). The future conditions
reported in (b) were extrapolated from the experimental data taken
at $R_0 = 20$~MHz and $\sqrt{s} = 8$~TeV. An overall factor
1.64 was applied to the expected rates to take into account
the increase of $\sqrt{s}$ from 8~TeV to 14~TeV.
The curve segments correspond to the interval of 
{\protect\matcalR$_{part}$} 
observed (a) and expected (b) in the different regions.}
\label{mu_efficiency}
\end{figure}

In Fig.~\ref{mu_efficiency} the muon detection efficiency for the seven
considered regions and readout
is reported as a function of $\matcalR_{part}$.
In Fig.~\ref{mu_efficiency}a the curve segments correspond 
to the effective $R^*$ intervals measured at 
the LHC $pp$ centre-of-mass energy,$\sqrt{s}=8$~TeV
with $R_0 = 20$~MHz and luminosity 
\hbox{$L = 4 \times 10^{32}$~cm$^{-2}$s$^{-1}$.}
In Fig.~\ref{mu_efficiency}b the curve segments are calculated
for $R_0 =40$~MHz and \hbox{$L = 2 \times 10^{33}$~cm$^{-2}$s$^{-1}$} 
and an overall factor 1.64 was applied to the expected rates to 
take into account the increase of $\sqrt{s}$ from 8~TeV to 14~TeV.
\vspace*{5truemm}

\begin{figure}[ht]
\begin{center}
\vspace*{0truemm}
\includegraphics[width=1.\textwidth]{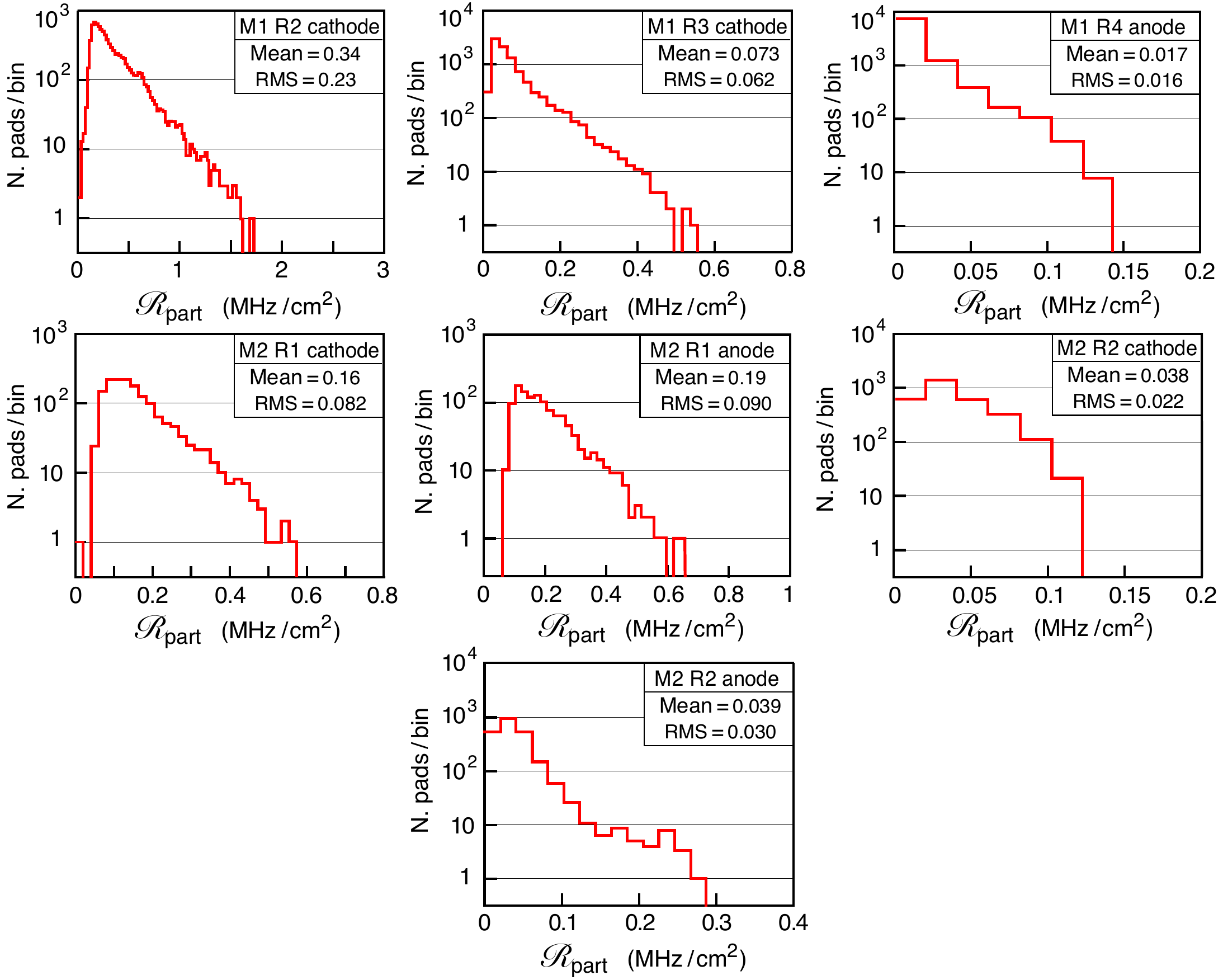}
\end{center}
\vspace*{-4truemm}
\caption{
Expected number of pads per bin of 
{\protect\matcalR$_{part}$} 
in the seven considered regions and readouts, at a luminosity 
of $2 \times 10^{33}$~cm$^{-2}$s$^{-1}$ and at $\sqrt{s}= 14$~TeV.
The distributions were extrapolated from the experimental data taken
at $R_0 = 20$~MHz and $\sqrt{s} = 8$~TeV. An overall factor
1.64 was applied to the expected rates to take into account
the increase of $\sqrt{s}$ from 8~TeV to 14~TeV.
}
\label{spettriR}
\end{figure}

Referring to the conditions of Fig.~\ref{mu_efficiency}b,
the number of pads irradiated with 
$\matcalR_{part}$~MHz/cm$^2$ is reported in Fig.~\ref{spettriR}.
The results presented in Fig.~\ref{mu_efficiency}b and in Fig.~\ref{spettriR}
show that in the future running conditions the inefficiency due to dead-time
would be a limiting factor to the operation of MWPCs belonging to
all the regions of station M1 and to the inner region of station M2.
Unacceptably poor FE efficiencies are expected in a relatively small zone
of these regions, near the beam pipe, where the rate of background (and of
good muons) is higher.
This effect contributed to the decision \cite{TDR} of 
removing station M1 in the upgraded detector for future 
high-luminosity runs at 14 TeV.
Regarding the inner part of M2, the effect of adding an 
additional shielding around the beam pipe has been studied \cite{TDR}
and other interventions on the detector
configuration are under investigation.
\vfill \eject
 
\section{Conclusions}

The dead-time $\delta_c$ of the front-end electronics of stations 
M1 and M2 of the LHCb muon detector
was determined from the experimental background 
rates measured at different luminosities by free-running 
front-end scalers.
The values obtained range  between $\sim 70$~ns and $\sim 100$~ns, with
the dead-time of the cathode pad readout being systematically larger than
that of the anode pads. These results allow to determine the  
muon detection inefficiency of a front-end channel due to the dead-time, 
which is the largest contribution to the detector inefficiency in station
M1 and in the inner regions of M2.
The muon detection inefficiency  was evaluated as a function of
the background rate per unit area of the pad for the LHCb Run1  bunch-crossing 
rate of 20~MHz and for the future condition when LHCb will run 
at a bunch-crossing rate of 40 MHz, at $\sqrt{s} = 14$~TeV and at a
luminosity of $2 \times 10^{33}$~cm$^{-2}$s$^{-1}$.
The large front-end inefficiencies expected in the future conditions
contributed to the decision of eliminating the station M1 and inserting an
additional shielding around the beam pipe upstream of the inner region of
station M2.

\appendix
\section{Dead-time effect on background counting and muon detection}

In the free-running front-end counters the hit of a particle generated
in the current BC can be lost because of the dead-time generated by a particle 
belonging to a preceding BC or by a particle belonging to the current BC.
When the luminosity is decreased, the time interval between
two BCs with interacting protons increases so that the 
counting losses due to preceeding BCs become negligible when $L \to 0$. 
This is not true for the counting losses due to particles belonging 
to the current BC. In fact when $L \to 0$ the number of interactions in
the considered BC is $n_{int} = 1$ and the average number of particles
hitting the pad is $\omega \neq 0$. 
Therefore when $L \to 0$ the fraction of lost particles 
$R_{lost}/R_{part} = (R_{part}-R^*)/R_{part}$ tends to a non-zero 
limit\footnote{If $\omega \ll 1$, which is the present case, this limit is
equal to $\omega /2$.}:
\begin{equation}
\label{eqappendix}
\lim_{L \to 0}\, \frac{R_{lost}}{R_{part}} = \lim_{L \to 0}\, 
\frac{R_{part}-R^*}{R_{part}} = 
\lim_{L \to 0}\, (\delta_{\,ef\!f}\, R^* ) \neq 0
\end{equation}
\vskip 1pt
\noindent
\begin{figure}[ht]
\begin{center}
\vspace*{-2truemm}
\includegraphics[width=.8\textwidth]{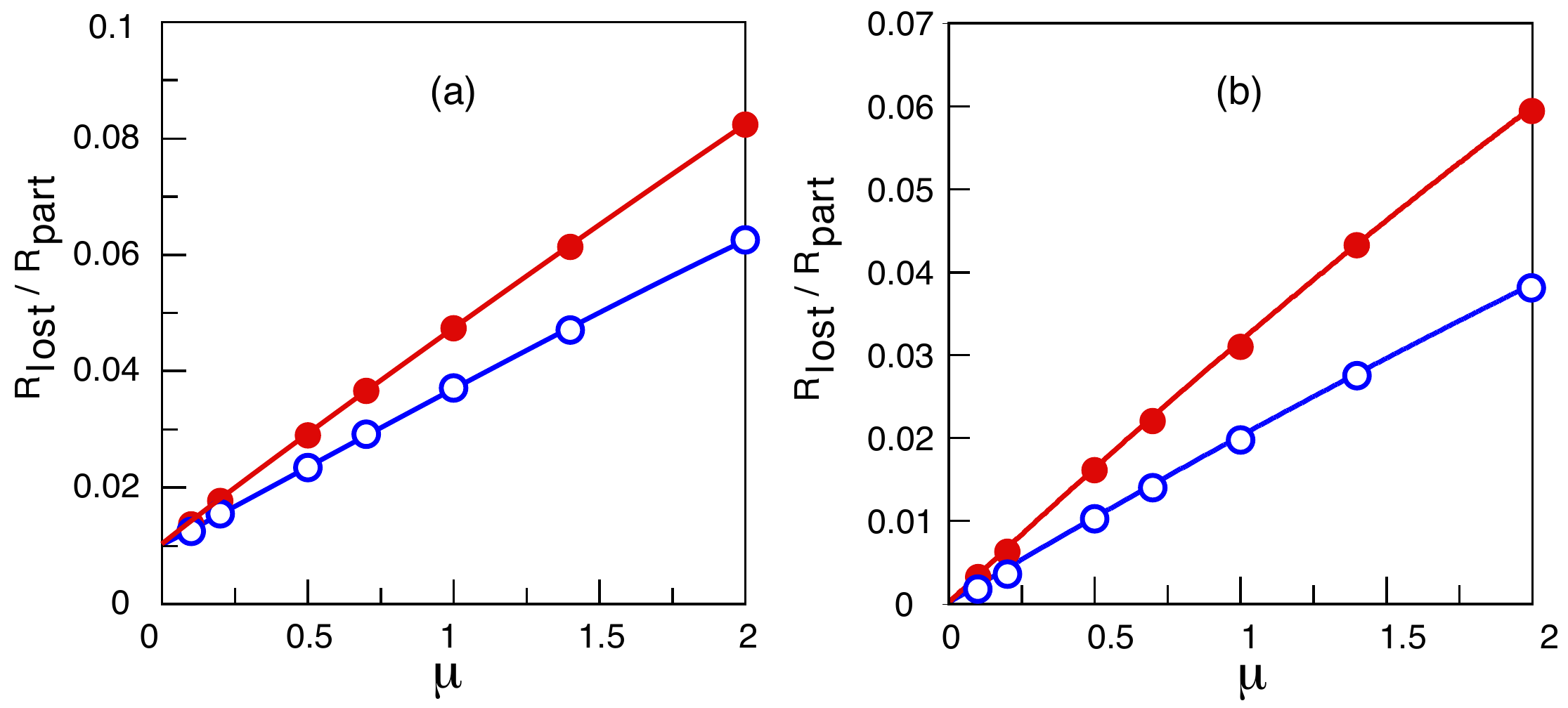}
\end{center}
\vspace*{-8truemm}
\caption{MC results on the ratio between the rate of lost particle and
the rate of hitting particles as a function of $\mu$ for $\omega=0.02$; 
(a) refers to background particles counted with the 
free-running scalers (see text) while (b) refers to muons 
counted in a 25~ns triggered gate (standard data taking situation). 
Open (full) points correspond to  $\delta_c = 70 \,(100)$~ns. 
}
\label{appendixnew}
\end{figure}

\noindent 
As shown in Fig.~\ref{appendixnew}a this effect is 
correctly predicted by the MC.
On the other hand when $\,L \to 0\,$ also $\,R^* \to 0\,$ 
so that the last equality 
in Eq.~\ref{eqappendix} implies that  
$\delta_{\,ef\!f} \to \infty$. This explains the behaviour of
$\delta_{\,ef\!f}$  shown in Fig.~\ref{deltaeff}a.

In the standard data taking the situation is quite different: 
the muon hit is acquired during a triggered gate of 25~ns. If in this gate 
the muon hit is cancelled by the dead-time of a background hit
generated and detected in the same BC, a hit is counted.
As a consequence, the fraction of muon hits lost depends only 
on the events belonging to the preceeding BC and goes to 0 when $L \to 0$.
Therefore $\delta_{\,ef\!f}^{(muon)}$ results, 
as expected, to be independent of
$\mu$. As shown in Fig.~\ref{appendixnew}b this
effect is correctly predicted by the MC.
\vfill \eject

\end{document}